\documentclass[a4paper]{jpconf}
\usepackage{graphicx}
\begin{document}
\title{Status of the TREX-DM experiment at the Canfranc Underground Laboratory}

\author{F~Aznar, J~Castel, S~Cebri\'an\footnote{Attending speaker.}, I~Coarasa, T~Dafni, J~Gal\'an\footnote{Present address: INPAC and Department of Physics
and Astronomy, Shanghai Jiao Tong University, Shanghai Laboratory
for Particle Physics and Cosmology, 200240 Shanghai, China.},
JG~Garza, FJ~Iguaz\footnote{Present address: IRFU, CEA, Universit\'e
Paris-Saclay, F-91191 Gif-sur-Yvette, France.}, IG~Irastorza,
G~Luz\'o´n, H~Mirallas, A~Ortiz de Sol\'orzano, E~Ruiz-Ch\'oliz,
JA~Villar\footnote{Deceased.}}

\address{Laboratorio de F\'isica Nuclear y Astropart\'iculas, Universidad de Zaragoza, Calle Pedro Cerbuna, 12, 50009 Zaragoza, Spain \\
Laboratorio Subterr\'aneo de Canfranc, Paseo de los Ayerbe s/n, 22880 Canfranc Estaci\'on, Huesca, Spain}

\ead{scebrian@unizar.es}

\begin{abstract}
The TREX-DM experiment is conceived to look for low mass WIMPs by
means of a gas time projection chamber equipped with novel
micromegas readout planes at the Canfranc Underground Laboratory.
The detector can hold 20~l of pressurized gas up to 10~bar, which
corresponds to 0.30~kg of Ar, or alternatively, 0.16~kg of Ne. The
micromegas will be read with a self-triggered acquisition, allowing
for effective thresholds below 0.4~keV (electron equivalent). The
preliminary background model, following a complete material
screening program, points to levels of the order of 1-10~counts
keV$^{-1}$ kg$^{-1}$ d$^{-1}$ in the region of interest, making
TREX-DM competitive. The status of the commissioning, description of
the background model and the corresponding WIMP sensitivity will be
presented here.
\end{abstract}

\section{Introduction}

Looking for low mass WIMPs which could be pervading the galactic
dark halo requires the use of light elements as target and detectors
with very low energy threshold and very low radioactive background.
Gas Time Projection Chambers (TPCs) with micromegas planes have
excellent features to fulfill these requirements. TREX-DM (TPC for
Rare Event eXperiments-Dark Matter) \cite{iguaz16,irastorza16} is a
micromegas-read High Pressure TPC for low mass WIMP searches using
Ar or Ne mixtures, not focused on directionality. The detector was
built and operated at surface in the University of Zaragoza as proof
of concept. The experiment has been approved by the Canfranc
Underground Laboratory (LSC) in Spain and is expected to be
installed underground by the end of 2017. The detector set-up and
performance are described in section~\ref{exp}, while the background
model developed and the corresponding sensitivity for WIMP direct
detection are discussed in sections~\ref{bac} and \ref{sen}.

\section{Detector set-up and performance}
\label{exp}

Micromegas are consolidated readout structures; a micro-mesh is
suspended over a pixelated anode plane, forming a thin gap where
charge amplification takes place. Detectable signals in the anode
and the mesh are generated. Different technologies have been built:
bulk micromegas have the readout plane and the mesh all in one and
microbulk micromegas are in addition more homogeneous and radiopure
\cite{microbulk}. They offer important advantages for rare event
detection: possibility of scaling-up, topological information to
discriminate backgrounds from the expected signal (just a few
microns track for dark matter, giving a point-like event) and low
intrinsic radioactivity as they are made out of kapton and copper,
potentially very clean. Indeed, after a first screening of
micromegas readouts using a germanium detector in Canfranc
\cite{mmradiopurity}, more sensitive measurements using the BiPo-3
detector \cite{bipo3} and a germanium detector with larger samples
are available now. The activity of the lower part of the $^{238}$U
and $^{232}$Th chains is below 0.1~$\mu$Bq/cm$^{2}$ \cite{jcapdbd}.
An activity of (3.45$\pm$0.40)~$\mu$Bq/cm$^{2}$ of $^{40}$K has been
quantified, which seems to be related to the production of holes by
kapton etching using a potassium compound.

\begin{figure}
\begin{center}
\begin{minipage}{18pc}
\includegraphics[width=0.8\textwidth]{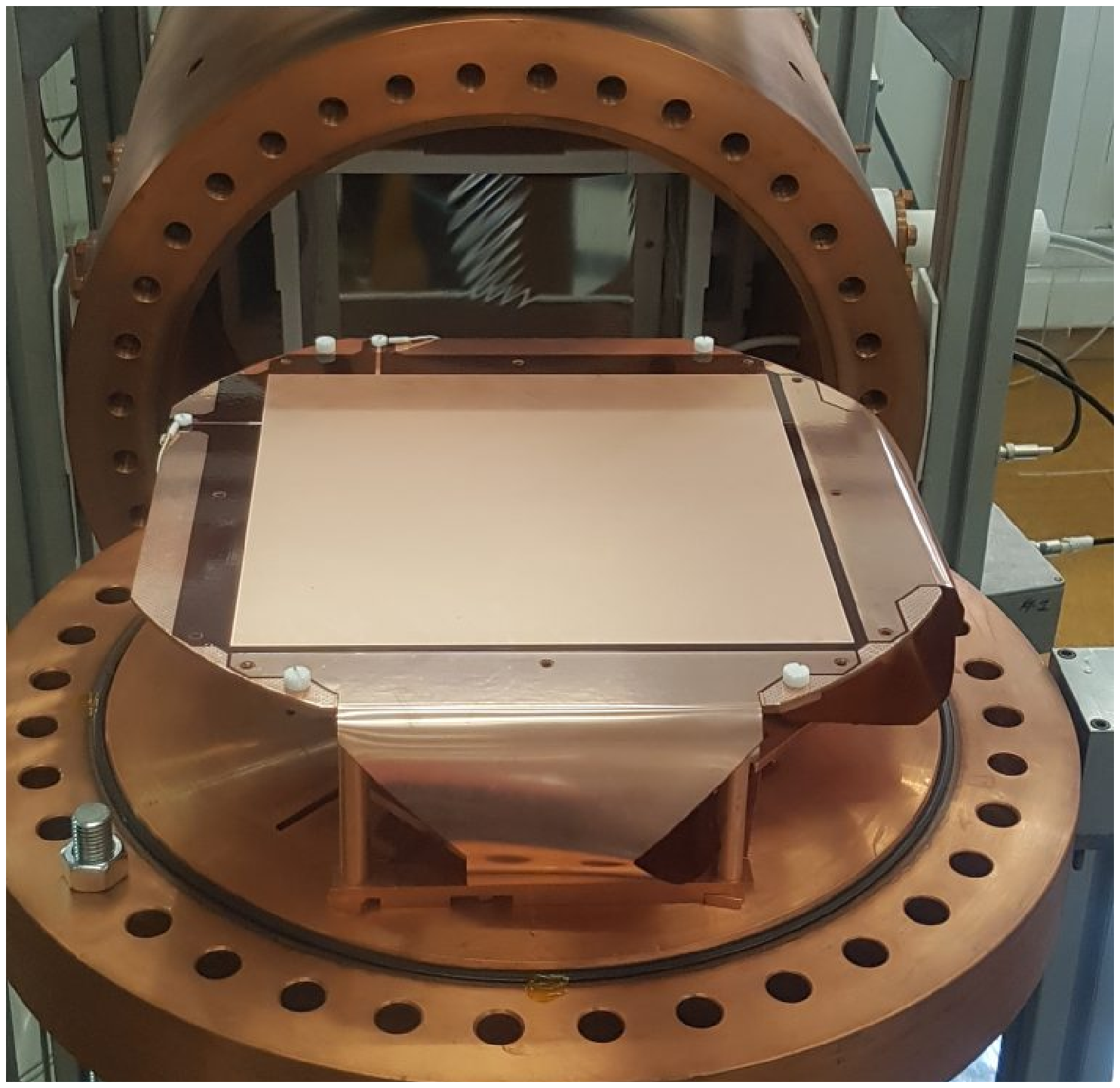}
\caption{\label{picture} Microbulk micromegas of TREX-DM (the
largest ever fabricated) being tested at the University of
Zaragoza.}
\end{minipage}\hspace{1pc}
\begin{minipage}{18pc}
\includegraphics[width=\textwidth]{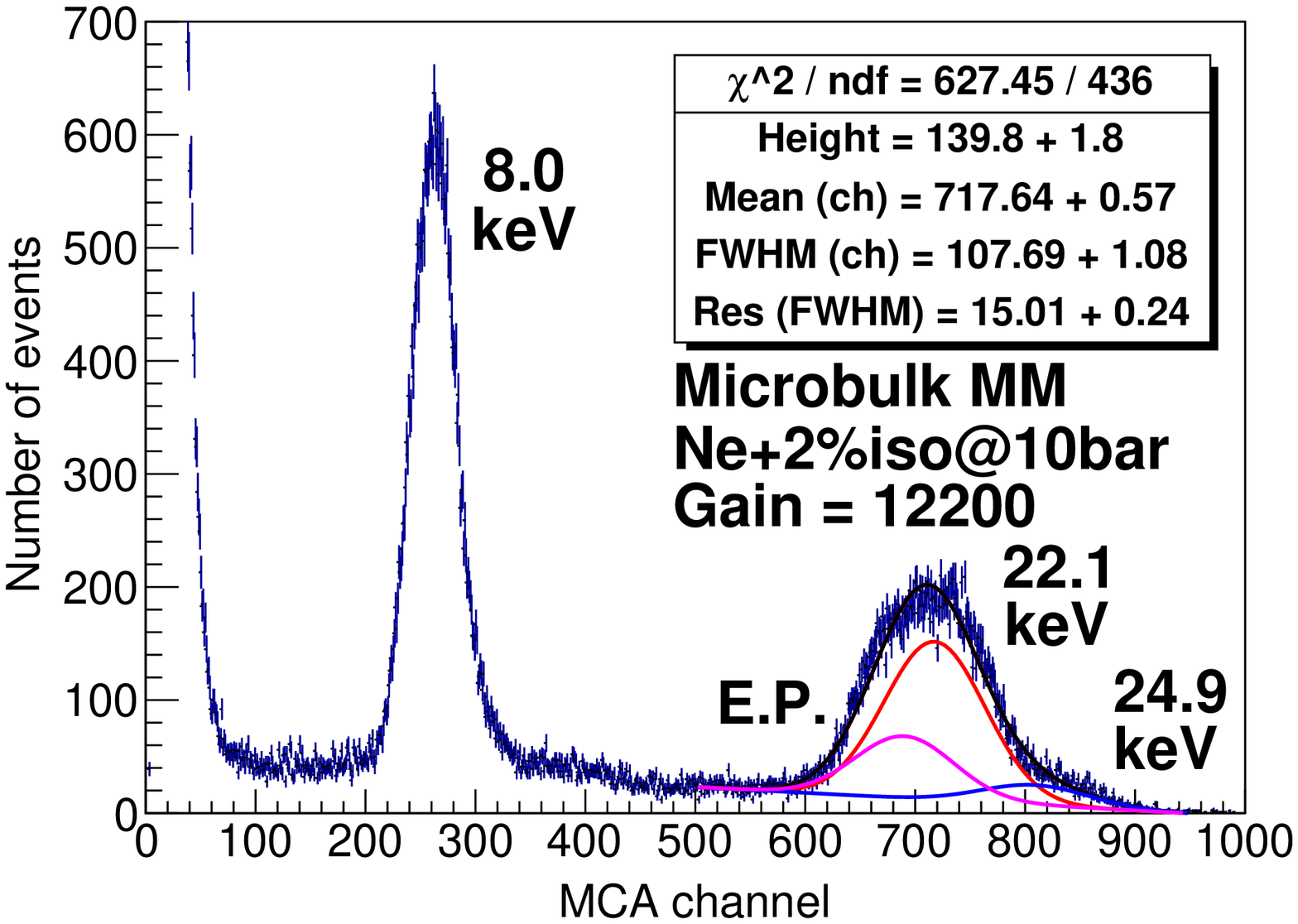}
\caption{\label{spcs} Energy spectrum obtained in the
characterization of microbulk micromegas with Ne using a $^{109}$Cd
source. The results of a multi-Gaussian fit for estimating the FWHM
at 22.1~keV are shown.}
\end{minipage}
\end{center}
\end{figure}

The TREX-DM detector, as built and operated at University of
Zaragoza, is described in detail in \cite{iguaz16}. Two active
volumes (19$\times$25$\times$25 cm$^{3}$ each) are separated by a
central cathode made of mylar inside a copper pressure vessel. The
field cage, made of kapton and copper, is covered by teflon. Two
bulk micromegas readouts were installed at the anode planes. Signals
were extracted by flat cables to the AFTER-based electronics. For
the set-up at hall A of LSC, the detector will be upgraded using new
more radiopure connectors, microbulk micromegas and an AGET-based
DAQ system; a complete shielding consisting of 5~cm of copper, 20~cm
of low activity lead, 40~cm of neutron moderator is being installed
and there will be a Rn-free atmosphere inside shielding. Signals
from 256$\times$256 strips ($\sim$1~mm pitch) will be digitized for
tracking and additional energy spectra are taken from the mesh. The
microbulk micromegas of TREX-DM are the largest area single
microbulk readout ever produced so far, with an active area of
25$\times$25~cm$^{2}$. After fabrication at CERN, they are being
tested inside the TREX-DM detector for the first time (see
figure~\ref{picture}).

First results from the commissioning phase of TREX-DM on surface
were shown in \cite{iguaz16}. More recently, microbulk micromegas
have been characterized in Ar+1\%iC$_{4}$H$_{10}$ and
Ne+2\%iC$_{4}$H$_{10}$ mixtures at 1-10~bar using a $^{109}$Cd
source. Energy resolution has shown some degradation with pressure,
being the FWHM at 10~bar 16(15)\% for Ar(Ne) at 22.1~keV (see
figure~\ref{spcs}). An excellent behavior has been registered for
gain, with maximum values above 10$^{3}$(10$^{4}$) in Ar(Ne) for all
pressures, which is very important for achieving low energy
thresholds. In principle, a very low threshold is possible thanks to
the intrinsic amplification in gas. In practice, the readout area,
the sensor capacitance and the electronic noise set the threshold. A
value of 0.45~keV$_{ee}$\footnote{Electron equivalent energy.} has
been currently achieved in a CAST-like detector using AFTER
electronics \cite{cast}. In the Zaragoza set-up, the trigger was
limited by the mesh channel noise level; therefore, trigger is going
to be obtained from low capacitance strips, using the AGET
electronics. In this way, the TREX-DM nominal (conservative) aim for
effective threshold is 100~eV$_{ee}$ (400~eV$_{ee}$).

\section{Background model}
\label{bac}

\begin{figure}
\begin{center}
\begin{minipage}{20pc} 
\includegraphics[width=\textwidth]{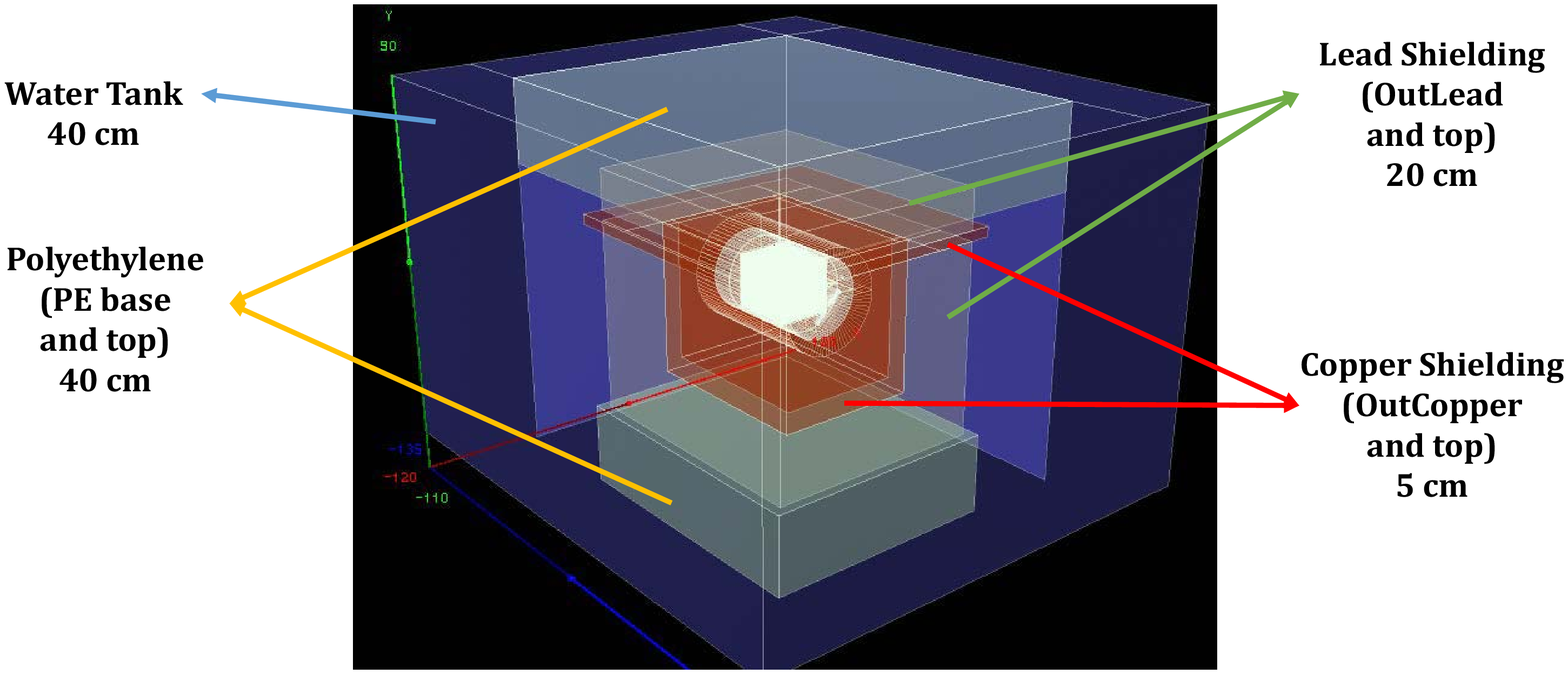} 
\caption{\label{geometry} TREX-DM geometry implemented in Geant4
simulations.}
\end{minipage}\hspace{1pc}
\begin{minipage}{16pc} 
\includegraphics[width=\textwidth]{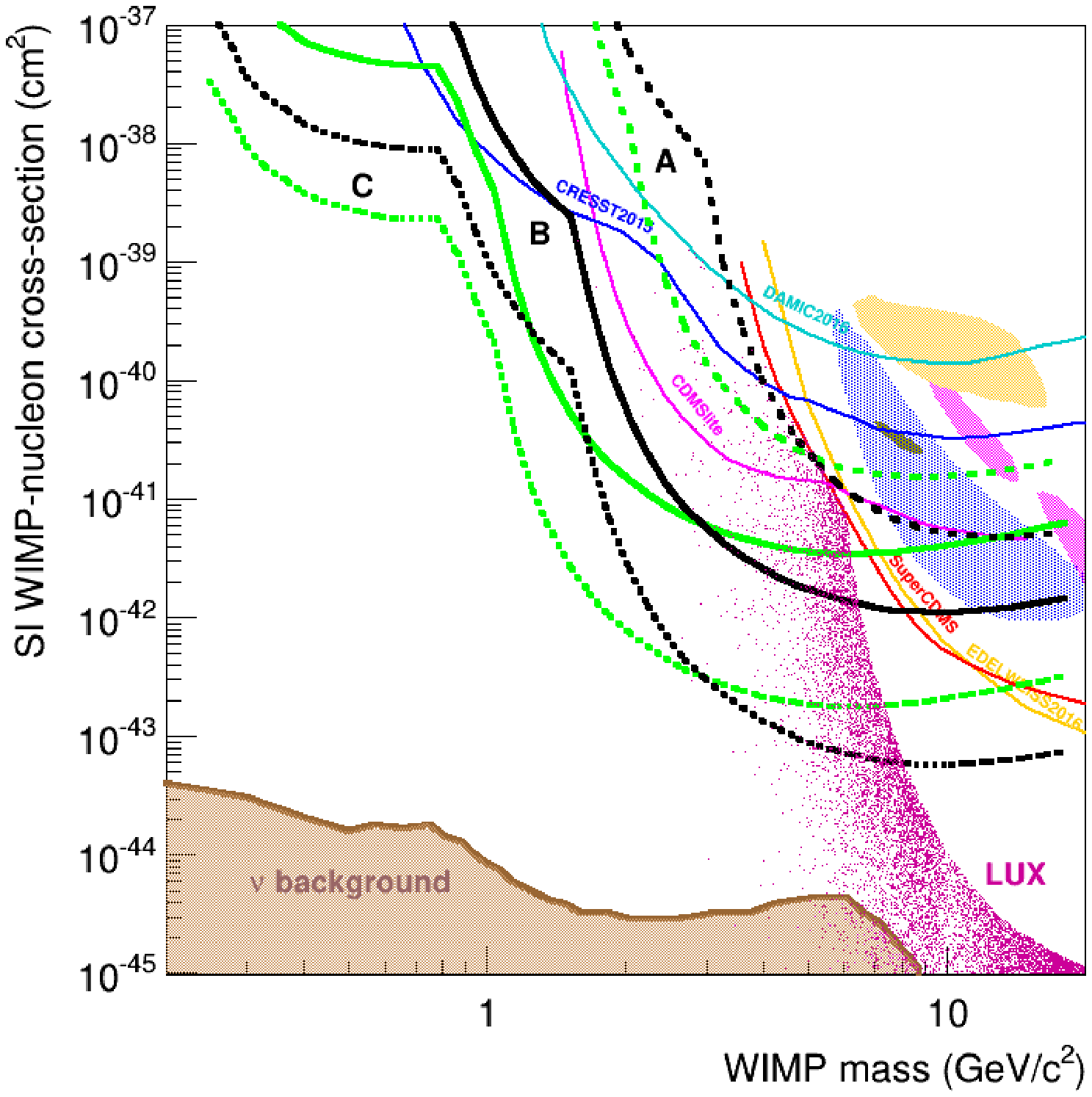} 
\caption{\label{exclusion} 90\% C.L. sensitivity of TREX-DM under
different conditions (see table~\ref{scenarios}) for
Ar+1\%iC$_{4}$H$_{10}$ (black lines) and Ne+2\%iC$_{4}$H$_{10}$
(green lines).}
\end{minipage}
\end{center}
\end{figure}

Ultra-low background conditions are a must in the direct detection
of WIMPs. An exhaustive material screening campaign underway for
several years has allowed to design and construct the detector and
shielding according to the radiopurity specifications
\cite{iguaz16,jcapdbd,jinstpaquito}; it is based on germanium gamma
spectrometry in Canfranc complemented by other techniques (GDMS,
ICPMS and BiPo-3 measurements). A preliminary background model of
TREX-DM for operation at LSC was presented in \cite{iguaz16} and is
being completed, including as inputs the material activity from the
screening program together with the measured fluxes of environmental
backgrounds at LSC (gamma-rays, neutrons and muons). Simulations of
the detector response are based on Geant4 (for physical processes)
and the custom-made REST code (for electron generation in gas,
diffusion effects, charge amplification at micromegas, signal
generation and analysis to select point-like events). A detailed
geometry of the set-up including shielding has been implemented (see
figure~\ref{geometry}) considering Ar and Ne mixtures at 10~bar.

Table~\ref{ratesin} presents the background rates from 2 to
7~keV$_{ee}$\footnote{This energy range corresponds to 5.2-16.3~keV
for Ar and 5.5-17.1~keV for Ne for nuclear recoils, following common
parameterizations of the quenching factor \cite{iguaz16}.} due to
primordial or cosmogenic activity in components inside or close to
the vessel. The largest contribution comes from the copper vessel,
cosmogenically activated after being a few years at sea level; an
activity of (0.24$\pm$0.05)~mBq/kg of $^{60}$Co was quantified in a
dedicated germanium measurement and included in the model. This
important contribution could be suppressed by constructing a new
vessel. The measured $^{40}$K activity in the micromegas readout
(see section~\ref{exp}) gives also a significant rate; new chemical
treatments are being analyzed to reduce this activity. The use of
underground argon has been assumed, considering the $^{39}$Ar
activity measured by DarkSide \cite{argon}. The total expected
background level is around 5(6)~counts keV$^{-1}$ kg$^{-1}$ d$^{-1}$
for Ar(Ne). Table~\ref{ratesout} shows the same background rates but
from activity in components outside the vessel and backgrounds at
the laboratory. The effect of radon-induced activity on copper
surfaces has been assessed, considering the deduced limit
($<$0.32~mBq/cm$^{2}$ of $^{210}$Pb) from a direct germanium
measurement on exposed copper. The contribution from muons and
environmental neutrons is under control in the simulated conditions.
All in all, the TREX-DM expected background is between 1 and
10~counts keV$^{-1}$ kg$^{-1}$ d$^{-1}$.

\begin{table}
\caption{\label{ratesin} Background rates (in counts keV$^{-1}$
kg$^{-1}$ d$^{-1}$) expected in 2-7~keV$_{ee}$ from activity in
components inside or close to the vessel using Ar or Ne mixtures in
TREX-DM. Isotopes giving the dominant contribution are indicated in
the last column.}
\begin{center}
\begin{tabular}{llll}
\br Component  &   Argon &  Neon    &  Main contribution \\ \mr
Vessel (primordial) & $<$0.088  &  $<$0.104  & $^{238}$U \\
Vessel (cosmogenic) & 1.25 & 1.50  &  $^{60}$Co \\
Copper Boxes (primordial) & $<$0.026 & $<$0.034 & $^{238}$U \\
Copper Boxes (cosmogenic) & 0.034 & 0.046 &  $^{60}$Co \\
Field Cage (PTFE) &  $<$0.033 & $<$0.051  & $^{238}$U \\
Field Cage (resistors) &  $<$0.35 &  $<$0.63 & $^{238}$U \\
Field Cage (kapton-Cu PCB)&  $<$1.06 &  $<$1.81 & $^{238}$U \\
Field Cage (cable)  &  $<$0.028 &  $<$0.052 & $^{238}$U \\
Cathode (copper) &  $<$0.0081  &  $<$0.012 & $^{238}$U, $^{40}$K  \\
Cathode (PTFE) &  $<$0.064  &  $<$0.085 & $^{238}$U  \\
Readout Planes &  $<$1.24 &  $<$1.14 & $^{40}$K  \\
Flat Cables &  $<$0.0097 &  $<$0.013 & $^{238}$U \\
Connectors &  $<$0.19 &  $<$0.24 & $^{238}$U \\
Epoxy &  $<$0.0044 &  $<$0.0056 & $^{232}$Th \\
Mesh Cable &  $<$6.1$\times$10$^{-4}$ & $<$7.7$\times$10$^{-4}$ & $^{238}$U \\
Other PTFE Components &  $<$0.017 &  $<$0.026 & $^{238}$U \\
Target & 0.15 & & $^{39}$Ar \\ \mr
Total & $<$4.6 & $<$5.8 & \\
\br
\end{tabular}
\end{center}
\end{table}

\begin{table}
\caption{\label{ratesout} Background rates (in counts keV$^{-1}$
kg$^{-1}$ d$^{-1}$) expected in 2-7~keV$_{ee}$ from activity in
components outside the vessel and backgrounds at LSC using Ar or Ne
mixtures in TREX-DM. Contributions marked with (*) are 90\% C.L.
limits when no event was registered in preliminary simulations.}
\begin{center}
\begin{tabular}{lll}
\br Component  &   Argon &  Neon    \\ \mr
Neutrons at LSC & (2.52$\pm$0.22)$\times$10$^{-2}$ & (7.06$\pm$0.61)$\times$10$^{-2}$ \\
Neutrons from $^{238}$U fission in Pb & (5.82$\pm$0.39)$\times$10$^{-5}$ & (1.094$\pm$0.074)$\times$10$^{-4}$  \\
Neutrons from $^{238}$U fission in Cu &  $<$2.1$\times$10$^{-6}$  & $<$4.1$\times$10$^{-6}$  \\
Muons (+ muon-induced neutrons) & 0.205$\pm$0.021  & 0.336$\pm$0.034 \\
$^{210}$Pb in Pb shielding (*) &  $<$0.12 & \\
Surface $^{210}$Pb on Cu vessel & $<$3.5$\times$10$^{-3}$ &  $<$6.2$\times$10$^{-3}$  \\
Surface $^{210}$Pb on Cu shielding (*) &  $<$0.025 & $<$0.034\\
Cosmogenic $^{60}$Co in Cu shielding & 0.0250$\pm$0.0018 & 0.0288$\pm$0.0020 \\
$^{222}$Rn in air & 0.1495$\pm$0.0024 & 0.0841$\pm$0.0013 \\
External gammas from $^{232}$Th  (*) & $<$9.9 & \\
External gammas from $^{238}$U (*) & $<$18 &  \\
External gammas from $^{40}$K (*) &  $<$27 & \\ \br
\end{tabular}
\end{center}
\end{table}

\section{Sensitivity prospects}
\label{sen}

Figure~\ref{exclusion} presents the attainable exclusion plots (90\%
C.L.) in the direct detection of WIMPs, for both Ar and Ne-based gas
mixtures at 10~bar, obtained assuming spin independent (SI)
interaction and standard values of the WIMP halo model and
astrophysical parameters. Three different scenarios for flat-shaped
background, energy threshold and exposure have been considered (see
table~\ref{scenarios}). A data-taking campaign of approximately
three years is foreseen, starting with Ne with the option to change
to depleted Ar. TREX-DM has a good potential to be sensitive to low
mass WIMPs beyond current bounds even at the scale of current
detector.

\begin{table}
\caption{\label{scenarios} Conditions assumed in the calculations of
the TREX-DM sensitivity shown in figure~\ref{exclusion}.}
\begin{center}
\begin{tabular}{llll}
\br & A  &   B &  C   \\ \mr
Background level (counts keV$^{-1}$ kg$^{-1}$ d$^{-1}$) &  10&  1 &  0.1 \\
Energy threshold (keV$_{ee}$) &   0.4 & 0.1 & 0.1 \\
Exposure (kg y) & 0.3 & 0.3 & 10 \\ \br
\end{tabular}
\end{center}
\end{table}

\section*{Acknowledgments} A few days after the TAUP2017 conference,
Professor J.A.~Villar passed away. Deeply in sorrow, we all thank
his dedicated work and kindness. This work has been financially
supported by the European Commission under the European Research
Council T-REX Starting Grant ref. ERC-2009-StG-240054 of the IDEAS
program of the 7th EU Framework Program and by the Spanish Ministry
of Economy and Competitiveness (MINECO) under Grants FPA2013-41085-P
and FPA2016-76978-C3-1-P. We acknowledge the technical support from
LSC and GIFNA staff.

\end{document}